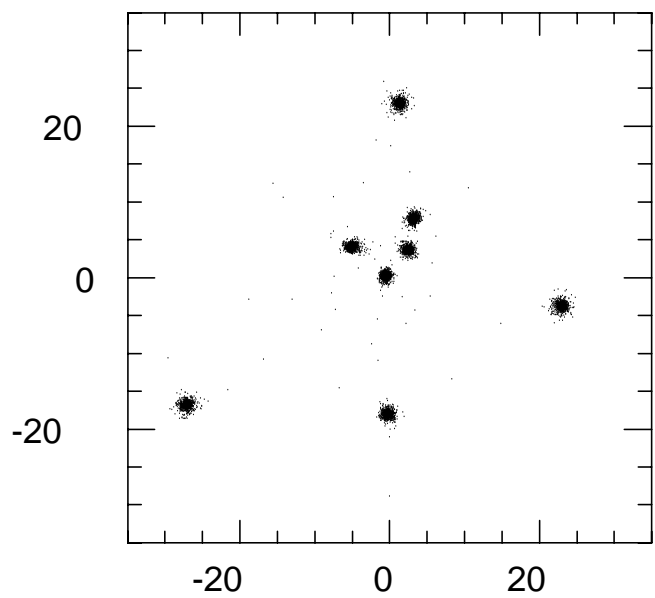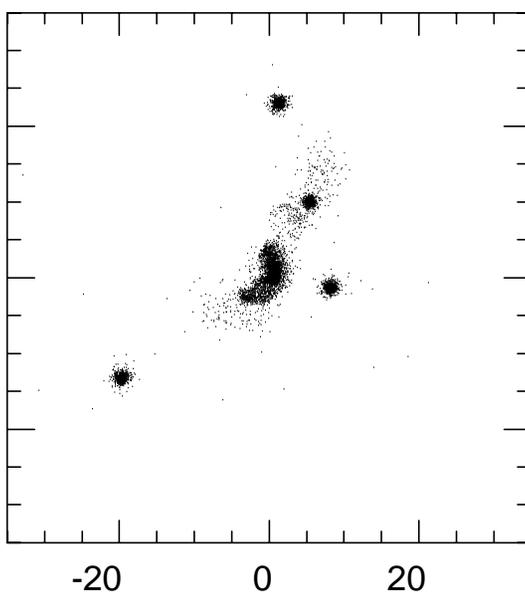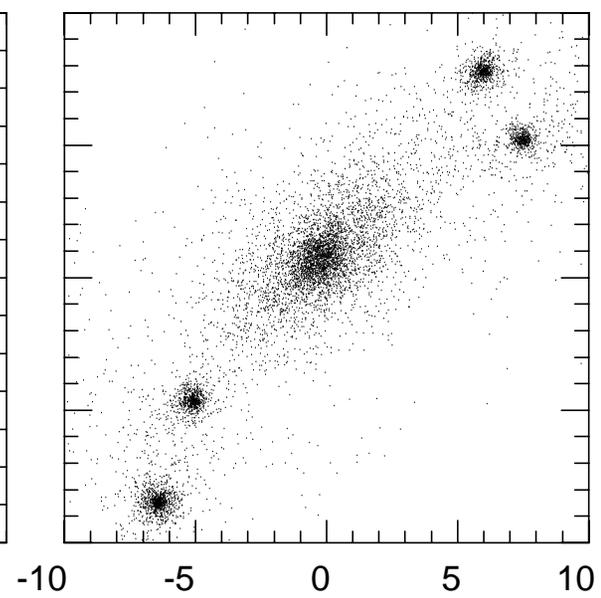

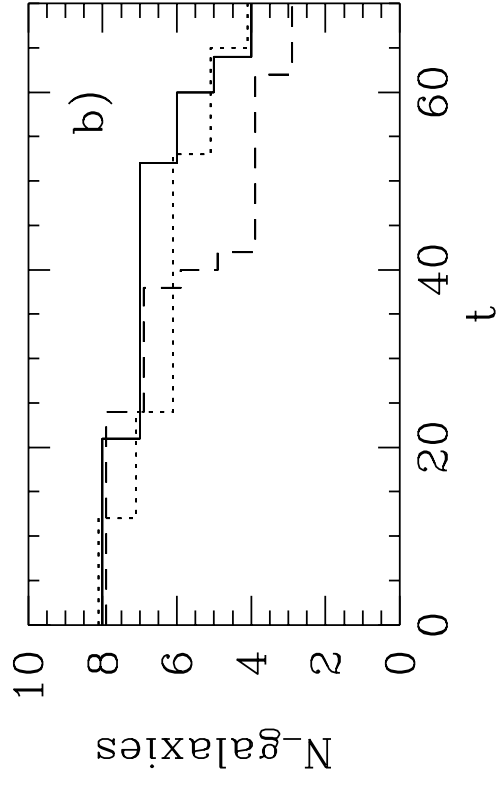
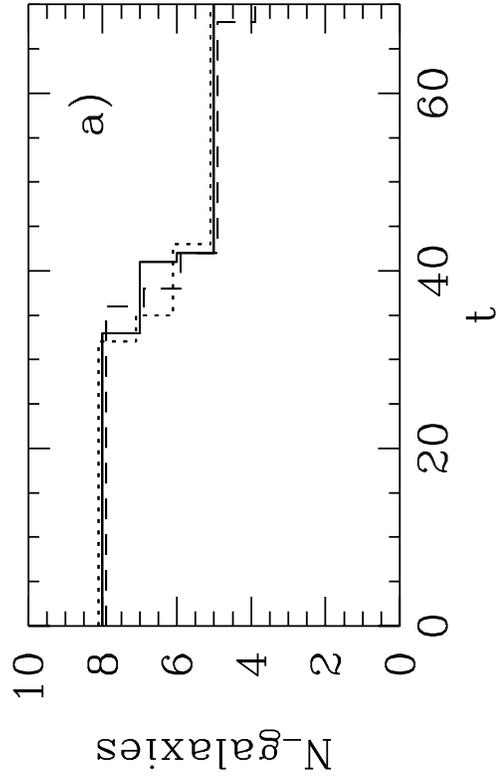

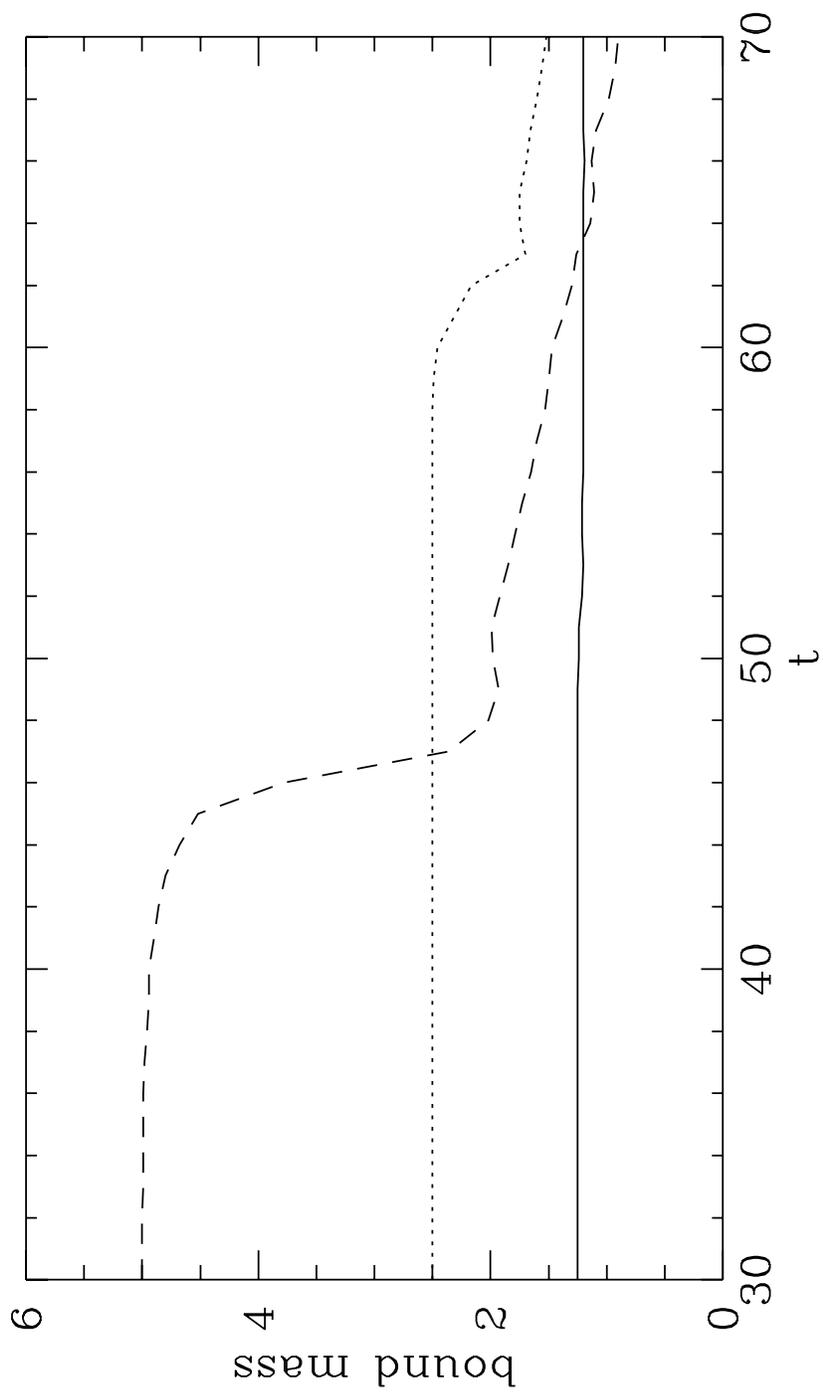

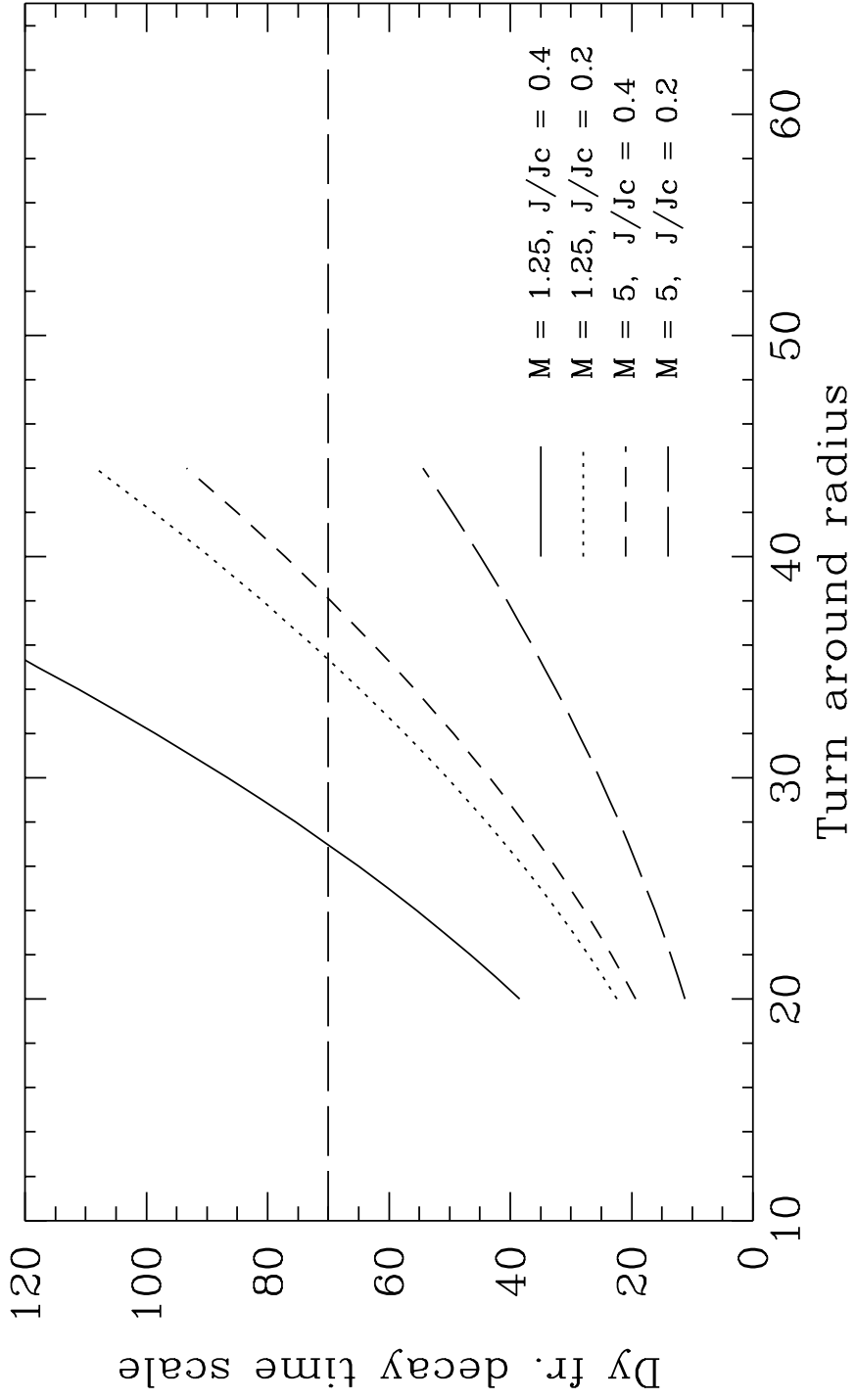

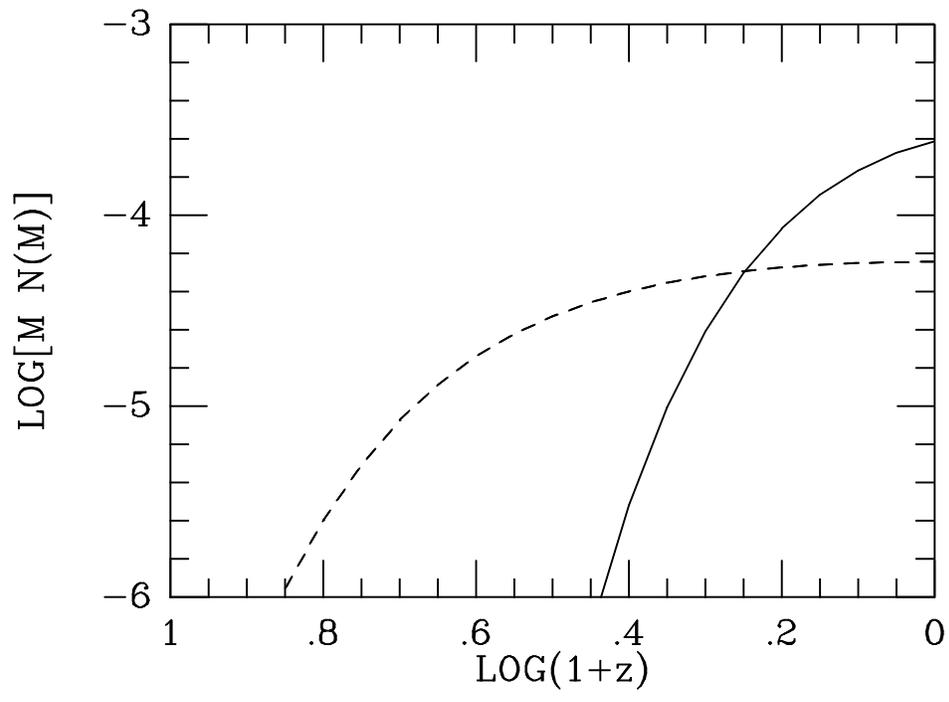

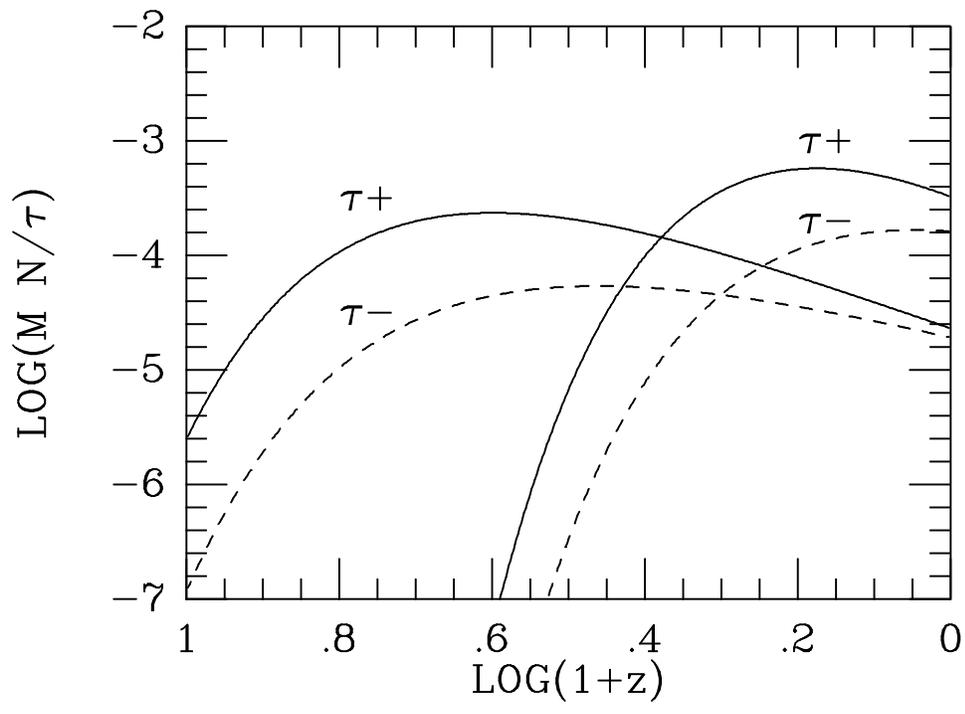

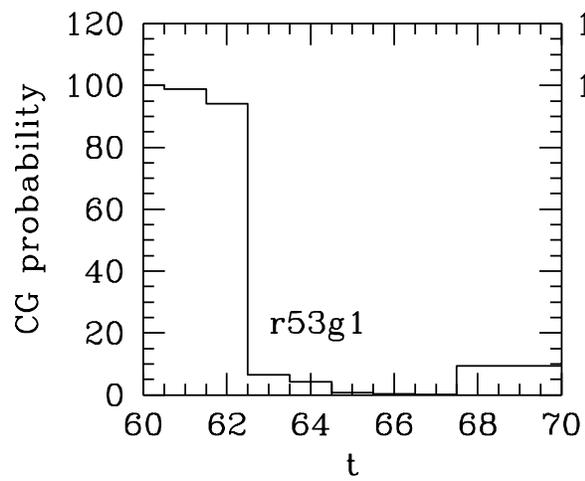
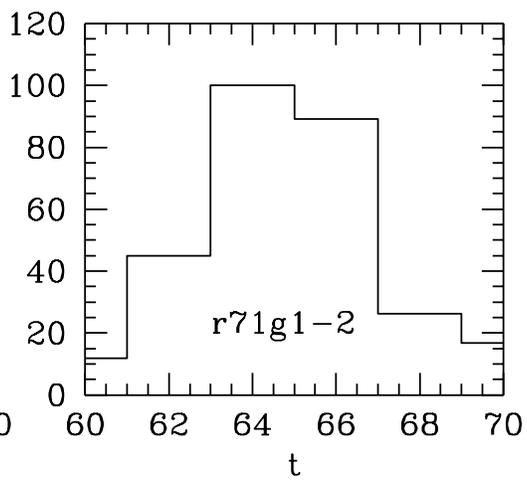
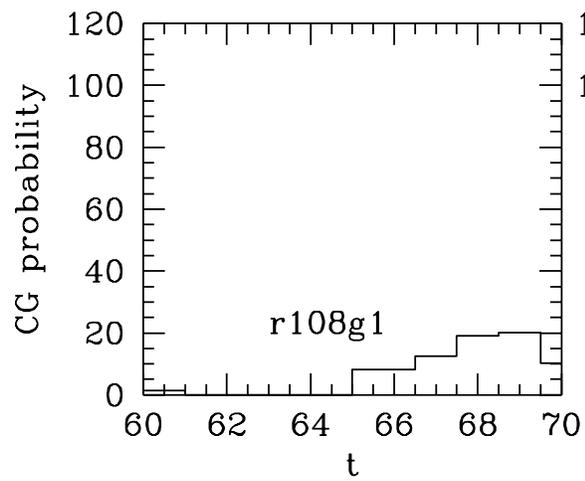
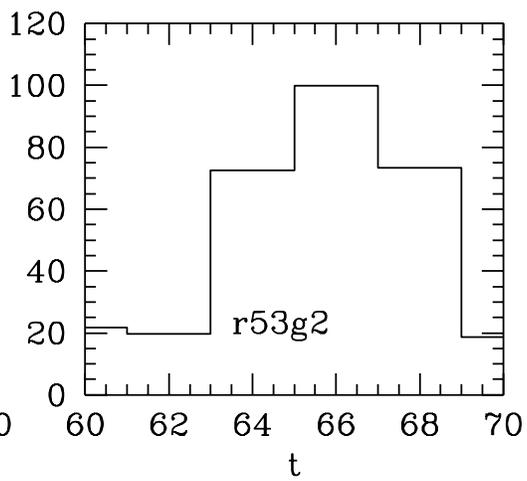

# Small Groups of Galaxies:
# a Clue to a Critical Universe

F. Governato[1], P. Tozzi

*II Università di Roma, via della Ricerca Scientifica 1, I-00133 Roma, Italy*

and

A. Cavaliere

*II Università di Roma, via della Ricerca Scientifica 1, I-00133 Roma, Italy*

## ABSTRACT

We study the formation and the subsequent evolution of galaxy groups with mass $2\ 10^{13} M_\odot$ in a critical universe, showing the importance of secondary infall for their dynamical evolution. From N-body simulations we find that merging is effective in building at least one massive central remnant in a few crossing times soon after the collapse of the central region of the group. Merging is only slightly accelerated if galaxies have massive halos, because the mass initially associated to the individual galaxies is soon tidally stripped. Stripping is particularly effective for infalling galaxies, which thus easily avoid merging with the central remnants. We find that, as a rule, merging is effectively terminated and the "merging runaway" ends when infall becomes dominant. We look for compact groups in our ensemble of simulations, and compare their statistical properties with Hickson's compact groups. We then discuss in terms of the Press & Schechter formalism the statistics of such groups and their evolution in different cosmological scenarios. Our results show that compact group formation is an ongoing and *frequent* process in a critical universe. In particular, our model reconciles the apparent contradiction between the observed absence of young–looking merger remnants and the high rate of strong galaxy interactions expected in compact groups. In open universes, instead, earlier formation of groups and suppression of secondary infall makes it more unlikely that compact groups survive as such until the present time. We conclude that the existence of dense and dynamically young groups of galaxies like HCGs concurs with other dynamical evidences pointing towards a high-density universe.

*Subject headings:* galaxies: interactions – clustering – cosmology: theory

---

[1] Also, Osservatorio Astronomico di Brera, Milano, Italy

## 1. Introduction

Groups constitute the most common galaxy associations, containing perhaps 50% of galaxies (Nolthenius & White, 1987) and so are an important laboratory to study the processes linked to galaxy formation and evolution. In particular, small groups of galaxies were the subject of several numerical studies (Carnevali, Cavaliere & Santangelo 1981; Barnes 1985), pointing out that galaxy merging tends to occur over a few crossing times of the group.

A much debated issue has been raised by the so called "compact groups" (hereafter CG, Hickson 1982, 1993; Prandoni, Iovino & MacGillivray 1994), i.e., small isolated galaxy associations containing at least 4 members. Many numerical works addressed the specific subject of small, dense groups ( Mamon 1987; Barnes 1989, Governato, Bhatia & Chincarini 1991). Their very existence appears to be in contrast with the notion that their high density and low velocity dispersion imply short merging timescales and hence rapid evolution into a single merged remnant (Sulentic 1987). Both their number density and the appearance of the member galaxies (large spiral fraction, lack of "blue" ellipticals) are apparently at variance with simple models of fast merging evolution (Zepf & Whitmore 1991; Zepf, Whitmore & Levison 1991; Mendes de Oliveira & Hickson 1994).

To overcome this problem, it has been suggested that a large fraction of the observed compact groups are not actually dense configuration, but rather chance superpositions of galaxies actually belonging to loose groups (Diaferio, Geller & Ramella 1994) or to filaments (Hernquist, Katz & Weinberg 1994, Ostriker, Lubin & Hernquist 1995 ). However, this interpretation does not account for the evidence of interactions among galaxies in CGs, many late type members showing distorted rotation curves and lack of neutral hydrogen (Rubin, Hunter & Ford 1991; Williams & Rood, 1987). In addition, a number of CGs show diffuse X-ray emission (Ponman & Bertram 1993; Ebeling, Voges & Boehringer 1993, Pildis, Bregman & Evrard, 1995; Saracco & Ciliegi 1995) proving their physical reality.

On the other hand, it has been pointed out that merging is delayed and the life of CGs prolonged if most of the group mass resides in a diffuse component so decreasing the cross section of the galaxies at a velocity dispersion fixed by the total mass (Cavaliere et al. 1983; Navarro & Mosconi 1987). However, without resorting to special initial conditions (high angular momentum, see Governato, Bhatia & Chincarini 1991) fast merging in a few crossing times can hardly be avoided, implying lifetimes for CGs less than a Hubble time. However, most of these simulations analyzed groups started either in a virial equilibrium state or with a simple top-hat distribution of galaxies sharing the Hubble expansion, so it is not clear how these results could be extrapolated to fully realistic conditions. What is still missing is a clear *evolutionary* path for these particularly dense galaxy associations, showing how and when they form in the general context of hierarchical clustering, and what is their final evolutionary stage.

We attack these problems on a broader front: first the detailed dynamical evolution of small groups is studied with N-body simulations; then the statistical number density of objects with a typical group mass ($\sim 10^{13} M_\odot$) is studied in terms of Press & Schechter (1974, hereafter PS) formalism. Finally, we discuss when and how compact groups form, and how the dynamical evolution of small groups is related to the issue of the density of the universe.

The paper is organized as follows: in §2 we discuss the N-body simulations. In §3 the statistics of small groups are studied with the PS formalism. §4 is devoted to CGs. In §5 the implications of our findings are summarized.

## 2. N-body simulations of small groups of galaxies

### 2.1. The codes used

The gravitational N–body simulations described in this work have been run using the TREECODE V3 kindly provided by Lars Hernquist (Hernquist & Katz 1989). In this code, the time step is adjusted individually for each particle; this results in shorter CPU times and in a more accurate integration in the denser regions of the systems where small time steps are needed. Tolerance was set to 0.7, quadrupole corrections were added, and energy is conserved to within 1% or better for all the simulations. Typically, trajectories of particles in regions with high velocity dispersion are integrated using many thousands of timesteps.

A number of simulations included a gaseous component. They were run using a gravitation + hydrodynamics code (treecode + SPH) kindly made available to us by Julio Navarro. A detailed description can be found in Navarro & White (1993).

## 2.2. The choice of the initial conditions

The general framework we consider in this paper is an Einstein–de Sitter universe with $h \equiv H_0/100 = 0.5$. The simulations start at $z = 5$, a time when galaxy halos have mostly collapsed and the bulk of star formation has taken place in bulges and elliptical galaxies (Renzini 1993).

Each simulated group is modeled as a spherical overdense region surrounded by a thick spherical shell with density equal to the mean density of the universe at $z = 5$. Both partake in the Hubble flow. The gravitational constant is set to unity and the total mass to 100, corresponding to $2 \times 10^{13} M_\odot$ with our preferred choice of units:

$$t = 173 \left(\frac{M}{2\ 10^{11} M_\odot}\right)^{-1/2} \left(\frac{l}{30\ \mathrm{kpc}}\right)^{3/2} \ \mathrm{Myr}, \quad (1)$$

where $t$ is the time unit, and $M$ and $l$ are the mass and length units.

Our initial conditions are such that the mass collapsed at the present time is the same for all our groups. Different initial overdensities $\delta_i$ imply different masses for the overdense region and for the external shell. The picture is similar to that described by the self–similar models of Bertschinger (1985). The main point is that for $\Omega_0 = 1$ the mass in the virialized part of the group is expected to grow as $t^{2/3}$ after the collapse of the initial perturbation, and to double between redshifts $z \approx 1$ and the present. This shows that groups described as simple top–hats without secondary infall are a poor description at least in a high density universe. In our model secondary infall has an increasing importance with increasing initial overdensity of the group, varying from 0% to 50% of the total mass at $z = 0$. When instead $\Omega_0 < 1$ holds, accretion ceases at $z \sim 1/\Omega_0$, so the simplifing assumption of no infall is legitimate.

As previously stated, the initial overdensity also defines the time of collapse for the central region. We choose four values for the initial overdensity (see Table 1), so that its collapse takes place at $z = 1.2, 0.68, 0.37$ and $0$, respectively. This means that we study groups similar as for virialized mass at the present time, but with a rather different history, in that their central parts virialized at very different epochs. The probability of such configurations is discussed in §3.

The group mass is divided between the galaxies and a diffuse component representing dark matter and smaller unresolved galaxies. The mass associated with individual galaxies for different models is 10%, 20% and 40% of the total mass, corresponding to $2.5\ 10^{11}$, $5\ 10^{11}$ and $10^{12}\ M_\odot$, respectively. The galaxy models used are described in detail in the next subsection.

Each run is identified by its initial overdensity (e.g., r71 for $\delta_i = 0.71$) and by the galaxy model used (g10, g20 and g40). An "s" suffix indicates runs with a gaseous component. Three runs (r71g20s, r71g10-2 and r71g40-2) were initialized with the same random seed, so galaxies start with the same spatial and velocity coordinates. In Table 1 the main properties of our group models are summarized.

Each simulation starts with 8 equal galaxies, roughly the number of expected bright galaxies in the range $2L_*$–$0.25L_*$ after the standard Schechter luminosity function for a group of total mass $2\ 10^{13} M_\odot$. Galaxies and background particles are randomly distributed inside the overdense region and the outer shell, in numbers proportional to the mass of each region.

The initial velocities for each galaxy and for the background particles are purely radial, following the Hubble flow with a correction factor proportional to the overdensity (Bertschinger 1985) $v(r) = Hr(1 - \delta/3)$, where $v(r)$ is the radial velocity and $H$ the Hubble constant at $z = 5$.

The model is clearly schematic as to initial shape and density profile, but adequate to our main scope here, that is, to study the effects of secondary infall on evolution and appearance of small galaxy groups.

## 2.3. Galaxy models

Our galaxy models include two components, one representing the stars (STARS) and the other the dark matter halo (DM). STARS are comprised of 320 particles and DM of 760. Including the background particles, each simulation uses about $2\ 10^4$ particles. The total STARS mass for each galaxy is fixed to $5\ 10^{10} M_\odot$. Different mass ratios of the dark–matter to the star component are used for three different models: 4, 9 and 19 (see Table 2).

STARS and DM components initially follow a King profile (King 1966), but with different values of the dimensionless central potential $W_o = 7$ and 1, respectively. The latter has been adopted because galaxies are to remain stable for a large number of crossing times against two–body scattering. This implies a rather large softening radius for DM particles (in our system of units $\epsilon = 0.4$; STARS have $\epsilon = 0.1$). With this softening, more centrally condensed halos would have been unresolved. The STARS component is kept fixed in mass and radius, DM halo sizes are scaled linearly with their mass, so more massive halos also have larger sizes and lower densities, conserving velocity dispersion and specific binding energy. Thus we explore a wide range of halos properties beyond the details of the chosen density distribution. A galaxy model is first built with a larger number of particles than used in the final simulations. STARS and DM are first evolved in the fixed external potential of the other component for many galaxy crossing times (defined as $CT = GM^{5/2}/|2E|^{3/2}$, where $E$ is the total energy) to ensure global equilibrium. Then the two components are added and let evolve together for a number of crossing times. Finally, DM and STARS particles are sorted by radius and evenly sampled to yield the actual galaxy model, whose parameters are listed in Table 2. These models when run in isolation are stable in mass and profile over times longer than those of the simulations. Morover, the softening and the number of particles used guarantee that spurious halo heating due to encounters with the more massive background's particles is negligible on the time scale of the simulations (Moore, Katz & Lake 1995)

### 2.4. The simulations

As the initial perturbation reaches turn–around and subsequently collapses, it undergoes violent relaxation and rapidly virializes. For each overdense region considered in isolation, its CT expressed in internal time units is given in Table 3 for different overdensity values. Galaxies inside the collapsed overdense region rapidly interact with each other, especially in runs r108, where the mean density is higher and the crossing time is accordingly shorter. They lose their halos and rapidly merge into a massive central remnant on a time scale of a few CTs. The evolution of r71g2-2, is shown in Figure 1.

The number of galaxies surviving in a group is shown in fig. 2 as a function of time for some representative simulations. No galaxy belonging to the initial perturbation in runs r108 is able to escape merging into a central merger, even for models with initially smaller halos; the process goes to completion in a few CTs. Such time scale is seen to apply for runs r71 and r53, where the process starts at lower redshifts; however, in such cases merging does not always involve all galaxies belonging to the initial perturbation.

Runs r35 are dominated by binary mergers during the long phase of expansion and recollapse, as seen in previous works (see Cavaliere *et al.* 1986).

In detail, galaxies with more massive and larger halos merge faster; but the difference is less than expected from previous works (see Bode, Cohn & Lugger 1993) which start from virialized initial conditions. This is because the collapse phase erases many initial details and the DM halos are cut by tidal stripping to about the same size and mass. As a consequence, with different halo models the merging times differ by less than a factor of two, this effect being also blurred by differences from one realization to another.

The group environment severely affects the mass associated to individual galaxies, since the tidal field is able to cut halo masses. Galaxies with large and massive halos (models g40) can lose a substantial fraction of their mass during their first pericentric passage. This process involves only the DM particles, while STARS particles corresponding to the visible part of galaxies are much less affected. This is because STARS are more centrally concentrated and bound. So STARS are removed only if galaxies interact with each other, and are about to merge.

Galaxies originally outside the overdense region reach the turn–around point, and fall toward the central region with large infall velocities. In most cases they never merge with the central remnant during the simulations. Fig. 3 shows the amount of mass loss for a number of infalling galaxies in different simulations.

These results may be understood in terms of two processes. As the infalling galaxies approach the central region their halo is stripped by the tidal field, thus reducing their cross section to a value nearly irrespective of the initial galaxy model. Then the orbital radii of the infalling galaxies decay due to dynamical friction. However, the decay is slow because of their small residual mass, and because of the large radius at which they turn–around.

A simple estimate of the time scale for the dynamical friction, $T_{fr}$, vs. the turn–around radius $r_{ta}$, is given by the classic formula (Binney & Tremaine 1987; Lacey & Cole 1993) which applies because the halo profile is nearly isothermal in its virialized part:

$$T_{fr} = 1.66 \frac{\sigma(r_{ta}/2)^2 \ (J/J_c)^{0.78}}{GM_g \ln \Lambda}, \qquad (2)$$

where $\sigma$ is the typical 3D velocity dispersion of our groups ($\sim 400$ Km/sec) and $J/J_c$ is the ratio between actual angular momentum and that of a circular orbit with the same energy. Fig. 4 shows decay times for different values of initial galaxy mass $M_g$, and for ratios $J/J_c$ as found in our simulations for infalling galaxies just before pericentric passage. We see that the decay times are long, especially for the lowest mass value (to which larger values converge by stripping, see fig. 3) and for the highest value of $J/J_c$, typical of infalling galaxies in our runs. These estimates agree with our numerical findings.

The main implication of this results is that in a critical universe the merging runaway (Cavaliere, Colafrancesco & Menci 1991) expected in small groups of galaxies is over by the time when secondary infall becomes important.

### 3. Small groups and the PS formalism

#### 3.1. Formation/destruction rates

To assess the statistical relevance of the findings in the previous section, two points must be addressed:
1) How representative are our initial conditions?
2) How common are small groups of galaxies at different epochs in different cosmologies?

To evaluate these points we first make use of the standard PS formula (see, e.g., Lacey & Cole 1993). It yields the number density $N(M,z)\, dM$ of high-contrast structures of a given mass, and its evolution with epoch.

The counting of such structures singles out volumes where the (linearly) evolved density contrast $\delta$ exceeds a nominal threshold $\delta_c$, which guarantees that the associated volume is collapsed into a virialized object. The canonical value $\delta_c = 1.68$ (see Peebles 1980) referring to the collapse of homogeneous spherical volumes is used throughout this paper.

The PS formula states:

$$N(M,z) = \sqrt{\frac{2}{\pi}} \frac{\rho\, b\, \delta_c\, a}{M_c^2(z)} m^{a-2}\, e^{-b^2 \delta_c^2\, m^{2a}/2}\ . \quad (3)$$

The normalization is such that at any $z$ all the available mass density $\rho$ resides in collapsed objects. In addition, $b^{-1}$ is the rms value of the linear fluctuation in the mass distribution on scales $8\, h^{-1}$ Mpc. Finally, we define $m \equiv M/M_c(z)$, where in a critical FRW universe $M_c(z) = M_{co}/(1+z)^{1/a}$ with index $a = (n+3)/6$ is connected to the Fourier power spectrum of the mass fluctuations taken of the form $P(k) \propto k^n$. In a low-density universe $M_c(z)$ evolves more slowly, corresponding to the approximate freeze out of the linear growth of the perturbations after $z \sim 1/\Omega_0$ (White & Rees 1978).

The main features of the PS mass function are: the power-law behavior $N(m) \propto m^{a-2}$ at the low mass end; the exponential cutoff at the high end. The (self-similar) evolution for decreasing redshift is described by the progression of the bright end to larger masses following the increase of $M_c(z)$, and by the mass-conserving decrease $1/M_c^2(z)$ of the normalization. These features correspond to reshuffling of small structures into larger ones.

This reshuffling can be successfully described recasting eq. (3) into the form of a rate equation (Cavaliere, Colafrancesco & Scaramella 1991):

$$\frac{dN}{dt} = \frac{N}{\tau_+} - \frac{N}{\tau_-}, \quad (4)$$

where in a critical universe

$$\tau_- = 3\, t/2 \qquad \tau_+ = \tau_-\, m^{-2a}/b^2 \delta_c^2. \quad (5)$$

The mass-independent timescale $\tau_-$ is associated to the destruction process by which groups are erased when enclosed into larger groups and clusters. In turn, $\tau_+$ is linked to the construction process, by which new groups form as the member galaxies are comprised by a collapsing perturbation. This scale is shortened at larger masses and with higher values of $b\delta_c$, implying a faster evolution. Build up of groups with mass $M_{gr} = 2\, 10^{13}\, M_\odot$ is dominant at epochs when $M_c(z) < M_{gr}$ holds, while destruction prevails thereafter.

#### 3.2. Application to small groups of galaxies

Now we focus on groups with mass $M_{gr} = 2\, 10^{13} M_\odot$. In terms of the PS formalism, we first evaluate the evolution of their number density as a function of redshift using eq. (3), and then discuss the average age of the population using eq. (4).

Most current fluctuation spectra are only gently curved on the scales of interest here, and so can be effectively approximated by simple power laws with an effective index $n$ slowly changing with scale.

The simplest CDM models fail to satisfy at the same time the normalizations constraints set at very large scales by COBE (Bennet et al. 1994), and at $8\,h^{-1}$ Mpc by the observed numbers of rich clusters (White, Efstathiou & Frenk 1993). We choose the power law approximation $P(k) \propto k^n$ which is not linked to a *specific* cosmological model. Then we choose to normalize our approximation $P(k) \propto k^n$ to the power spectrum using only the latter constraint, the closest to the scale of our interest.

We consider two FRW universes, one with $\Omega_0 = 1$ and the other with $\Omega_0 = 0.2$. We consider for the amplitude the values $b = 1.7$ for the former ("critical case") and $b = 1$ for the latter ("open case"), after Bahcall & Cen 1993, Henry & Arnaud 1992, and retain the usual value $\delta_c = 1.68$. For the spectral index we use a set of representative values: $n = -1.5, -2, -2.5$. For comparison, $n = -1.8$ is the value required for a CDM spectrum with $\Omega_0 = 1$, and $n = -2$ corresponds to CDM with $\Omega_0 = 0.2$ (Holtzmann 1989). Note that the cluster number density which sets our normalization depends only weakly on the index $n$ (White, Efstathiou & Frenk 1993).

In Fig. 5 we plot the number density vs redshift for the intermediate index value $n = -2$. The number density first rises rapidly on the time scale $\tau_+ = m^{-2a}\,3t/2$ when $M_c(z) \leq M_{gr}$ holds; then it saturates as destruction operates on the slow scale $\tau_- = 3t/2$. In the critical case the group aboundance rises continuously till the present epoch. In the open case, the number density of groups is nearly constant from $z \approx 3$ to the present.

The local group abundance depends only weakly on $b$ because the group mass $M_{gr}$ lies in the power law section of the mass function $N(M)$, well below the position of the exponential cutoff. For a low–density universe, $M_{gr}$ is closer to the cutoff and the group aboundance depends on $b$ to some extent. So the number densities predicted by PS are $\sim 3 \times 10^{-4}$ Mpc$^{-3}$ for $\Omega_0 = 1$ and about a factor of 5 lower for $\Omega_0 = 0.2$. The observed numbers (Bahcall & Cen 1993; Moore, Frenk & White 1993; Nolthenius 1993) discussed below, are intermediate, and do not discriminate between the two universes, considering the uncertainties.

Of greater interest here are the ages of the group population, as old groups are expected to be more dynamically evolved. We use the rate equation (4) to obtain the formation and the destruction rates $MN(M)/\tau_\pm$ for groups as a function of $z$ (see Fig.6).

The net number variation at a given mass in a time interval $\Delta t$ is given by the difference between the fraction of newly formed groups and that of old ones which are subsumed into more massive structures: $\Delta N/N = (\Delta N/N)_+ - (\Delta N/N)_- = \Delta t/\tau_+ - \Delta t/\tau_-$. So we evaluate separately the percentage of the new and the older population for the critical and open case, and for three values of the effective index $n$.

Focusing on objects born since $z = 0.3$ (an epoch actually reachable in deep surveys), in the critical case we obtain 34%, 56% and 79% for $n = -1.5, -2, -2.5$, respectively, relative to the local population $N(M_{gr}, 0)$. Analogously, we obtain 25%, 22% and 18% for the objects destroyed since $z = 0.3$. We note an efficient turnover of the population, especially in the number of newly formed objects.

With the open case we obtain 10%, 15% and 20% of objects born since $z = 0.3$, and 11% of objects destroyed since, nearly irrespective of the index $n$. In this case very few objects appear later than $z = 0.3$.

The time scale for galaxies in small groups to suffer extensive merging is estimated to be just a few CTs so it is interesting to express the age of the groups also in terms of their crossing times. Referring to the formation epoch $z_f$ when half of the present population is formed, we obtain $z_f = 0.43, 0.27, 0.16$ in the critical case, for $n = -1.5, -2, -2.5$, and $z_f = 1.4, 1, 0.8$ for the open case. For the age in terms of crossing time of an object formed at $z_f$, we obtain (conservative) ages in the range $1 - 2$ CTs for the critical case, and in the range $4 - 7$ CTs for the open one. We conclude that in open universes the population of groups is made of objects dynamically older than in the critical case.

We note that if the bias were kept at the value $b = 1$ even in the critical case (which would violate the observed number density of rich clusters), still the fraction of newly formed groups would differ considerably from the open case.

The main result of this section is that the population of groups differs *widely* in a critical and in a low density universes in terms of age. In the former a larger fraction of groups seen at present time were born at relatively recent redshifts. In the latter universe, on the other hand, most groups formed at relatively high redshifts ($z > 0.3$), and now their formation rate is low. These results depend only weakly on the spectral index $n$.

Finally, we evaluate the significance of the initial conditions used in our simulations. We evaluate the redshift for the central regions to collapse; specifically, we evaluate $z_{1/2}$, the redshift when half of the final mass of the group is enclosed into a single progenitor condensation. In terms of the conditional mass function $N(M, z|M_{gr}, 0)$ for the progenitors of a group with given collapsed mass $M_{gr}$ at redshift $z = 0$, following Bower 1991 and Lacey and Cole 1993 we set the condition:

$$\int_{M_{gr}/2}^{\infty} dM \; M \; N(M, z|M_{gr}, 0) = 1/2 , \quad (6)$$

and solve for $z_{1/2}$. In the critical case Bower 1991 gives the explicit solution:

$$z_{1/2} = \frac{0.97}{b\delta_c} \left(\frac{M_{gr}}{M_{co}}\right)^{-a} (2^{2a} - 1)^{1/2} . \quad (7)$$

For example, for $n < -1$, redshifts $z_{1/2} < 1.7/b$ are obtained, and for $n = -2$ the value is $z_{1/2} = 0.6/b$ (note that for more negative $n$ the evolution is faster, as it is for higher bias). This shows that runs with earlier collapse of the central region, like runs r108, r71 and r53 in our Table 1, are likely realizations for $-2 < n < -1.5$. Instead runs r35, i.e., top–hats collapsing at very low redshift, are less probable.

## 4. Compact groups of galaxies

### 4.1. Observed CGs

The defining criteria for Hickson's compact groups (HCGs, Hickson 1982) are:

   i) surface brightness, $> 26$ mag /arcsec$^2$;
   ii) membership, $N_{gal} \geq 4$;
   iii) magnitude concordance, $m_f - m_b < 3$;
   iv) isolation, $3 \; R_{gr} < R_{nn}$ .

Here $m_b$ and $m_f$ are the magnitudes of the brightest and faintest galaxy, $R_{gr}$ and $R_{nn}$ are the radius of the minimum circle enclosing the group and the distance to the nearest galaxy outside $R_{gr}$ satisfying iii).

Because dense groups constitute an environment where merging should be common, and their persistence to the present is sensitive to cosmology as discussed in §2 and §3, we undertook a thorough search for CG-like configurations in our N–body simulations, incorporating Hickson's criteria into our finding algorithm. The aim was to test the possibility of originating CGs in the framework of a critical universe.

### 4.2. CG occurrence in N–body simulations

Each simulation was first examined at many output times, and subjected to many (1000) random rotations, to compute the probability to observe a compact configuration meeting all Hickson's criteria.

It is not straightforward to establish the luminosity of each galaxy, as tidal stripping lowers the bound mass, while on the other hand interactions can substantially raise the luminosity of a given galaxy (Xu & Sulentic 1991). We assign to each galaxy a luminosity proportional to its initial STARS mass or, in case of a merger remnant, to the sum of the corresponding masses of the progenitor galaxies. We assume a ratio $M/L = 5$ $(M_\odot/L_\odot)$ in the R–band for the STARS particles, implying that each model galaxy has an initial luminosity of about $0.25L_*$ and the mass over light ratio M/L is 250 for whole group.

Is our "secondary infall model" able to generate HCGs? In fig. 7 we show for four simulations the probability vs. time for the group to be selected as a HCG between redshifts $0.1 - -0$, the range for the observed HCGs. For each realization the probability that the association meets Hickson's criteria oscillates with time, corresponding to physical events like a merging or the infall of one galaxy, or corresponding to geometrical circumstances. On average, however, *the majority of our simulated groups have a fairly high probability to appear as a HCG*. Integrating for $z < 0.1$, and averaging over our more probable realizations with the central overdensity collapsing at $z_{1/2} \approx 1 - 0.35$, the probability is about 25%. Most ($\sim 70\%$) of such compact configurations are real, and not superpositions of galaxies far out in the group. The ability of our model to create compact configurations similar to the observed ones suggests that secondary infall must be an important, and probably necessary ingredient to explain the existence of CGs. Moreover, our simulations indicate that HCG configurations are not isolated groups of galaxies that collapse and then merely survive merging. Rather, the compact appearance is due to secondary infall replenishing the group with new galaxies which, as described in the §2.4, rarely merge with the central remnant.

Keeping in mind the simplicity of our model, it is interesting to check its outcome as for the expected number density of CGs in a critical universe. As a first, crude approximation, this can be done on multiplying the number density of groups by the probability of observing a CG inside them. We use the local number density observed on the mass scale $M_{gr}$ and in the magnitude range $m_R = -21, -23$, which is of the order $3 \; 10^{-4}$ Mpc$^{-3}$ (Bahcall & Cen 1993; Moore,

Frenk & White 1993; Nolthenius 1993), roughly consistent with the PS estimate given in §3.2, thus the observed number density of groups turns out to be about 50–40 times larger than the observed density of HCGs ( Mendes de Oliveira & Hickson 1991).

To obtain an estimate of the *observed* number density of CGs, one has to consider first the probability of about 1/4, as said above, of actually seeing a compact group. To make a detailed comparison with the Hickson catalog, various selection effects should be considered. Superpositions of unrelated galaxies strongly diminish the observable number of CGs by violating the isolation criterion. Many are due to distant, unrelated background or foreground galaxies. Using a projected Poisson distribution we evaluate the probability of isolation $P_{is}$ (see Hickson, Kindl & Auman 1989) for a typical HCG of fixed radius $R = 0.1$ Mpc and brightest member magnitude $-21$, and find $P_{is} \simeq 0.3$ out to $z \approx 0.03$, where the majority of the HCGs are located (Hickson, Mendes de Oliveira & Palumbo 1992). At this point, our estimated number density is about three times the observed one, which constitutes a fair agreement given the catalog incompleteness and the simplicity of our model.

A more subtle source of superposition, which would lower the observed number of CGs by violating the isolation condition, is due to galaxies close in projection but not dynamically bound to the subset defining the HCG (see Ramella et al. 1994). The distribution of such galaxies, as they lie mostly on sheets and filaments, is obviously non–Poissonian and the importance of this effect is difficult to evaluate without resorting directly to cosmological simulations. It has already been shown (Hernquist, Katz & Weinberg 1994) that filamentary structures can produce fake CGs. Determining the actual importance of this source of contamination and so a precise estimate of the number density of CGs is beyond the scope of this work and needs to be evaluated from high resolution cosmological simulations, including galaxy formation and a full galaxy luminosity function.

Our simple estimates suggest that the secondary infall model in a critical universe *is able to predict a number density of CGS consistent with observations*. As previously discussed in §2 and §3, in an open universe groups lack effective secondary infall and are substantially older in terms of CTs. As a consequence the initial set of galaxies suffers fast merging on a time scale of just a few CTs, as suggested by previous works on small groups not including secondary infall.

These arguments strongly suggest that in a low density universe the number density of CGs should be much less than observed.

### 4.3. Observed and simulated CGs

A number of observed features of real CGs can be related with the results of our simulations, and used to test the secondary infall model.

We find crossing times of the simulated compact groups of order $10^{-2}/H_0$, quite in agreement with the observed ones (Hickson, Mendes de Oliveira & Palumbo 1992).

In our simulations, non–compact configurations, like those including all the galaxies in a group (often CGs are isolated quartets within a non compact quintet or a sextet), and especially groups collapsing at the present time (like all runs r35) have longer CTs, if observed between z = 0.1 and 0. We relate these with CTs measured for normal groups by, e.g.,Ramella, Geller & Huchra 1989, Nolthenius 1993; Moore, Frenk & White 1993).

Moreover, we find that surviving galaxies are segregated with respect to the dark matter background, so that estimates of the virial mass $M_{vir}$ from CG members are almost always smaller by about a factor of two than the total mass $M_{tot}$.

Among HCGs, many comprise very bright ellipticals (Mendes de Oliveira & Hickson 1991). These qualify as first candidates for merging remnants. But surprisingly these galaxies do *not* show any obvious sign of recent merging activity, like double nuclei, shells or blue colors (Zepf & Whitmore 1991). This apparent contradiction is just what is expected in our model: almost all of our simulated groups underwent a phase of strong interactions and merging long ago, during the collapse of the central regions at redshifts $z \approx 1-0.35$ for our more probable realizations. In the simulations, during the collapse phase one or two massive mergers are almost always formed. Thereafter merging phenomena are strongly suppressed during the secondary infall phase, so these galaxies at present time would not show anymore the observational features associated with merger remnants. This scenario is also supported by recent observations of ellipticals in CGs (Caon *et al.* 1994), showing that most elliptical suffered major mergers, but only 2 or more Gyrs ago.

A complementary problem is indicated by observations of the spirals in CGs. Many of them show signs of interactions compared to isolated spirals, namely peculiar rotation curves (Rubin, Hunter & Ford 1991), HI deficiency (Williams & Rood 1987) and low radio emission (Menon 1991). These signs induce expectation of at least some strong interactions, recent or still ongoing. But if the spirals are strongly interacting, where are the corresponding merging remnants? The results of our simulations suggest that most of such features seen at present time are due to the group tidal field and not to merging. With the plausible hypothesis that infalling galaxies are preferentially late type, tidal interaction with the group as a whole will perturb their internal density distribution. In addition, the intra–group medium, revealed by high resolution X–ray observations in a number of HCGs ( Pildis, Bregman & Evrard, 1995; Saracco & Ciliegi 1995), will tend to strip galaxies out of their HI content during the first infall.

To assess the importance of ram pressure we have run two additional simulations (r108g10s, r71g20s) using an N–body + SPH code to treat the gaseous component. The gas mass and the associated particle number were chosen to be equal to the total STARS component. Initially the gas was distributed like the diffuse DM background, at a temperature $10^4$ K.

Shock treatment was included, but no gas cooling was allowed. At the end of the simulations the gas density profile was measured, finding in all three runs $n \sim 3\ 10^{-4}$ electrons/cm$^3$ within a radius of 100 kpc. At this radius, the infall velocity reaches 800 km/s. The stripping efficiency is estimated (see Sarazin 1988) on the basis of the ram pressure $\rho v^2$, which turns out to be 0.1 − 0.3 ($\Omega_b/0.02$) relative to that typical of the central regions of rich clusters. This preliminary result suggests that ram pressure stripping may be effective, yet not overwhelming, for the outer regions of the infalling spirals.

It must be noted that systematic HI deficiency in spirals of a CG (due to either tidal or ram–pressure stripping) constitutes a strong argument against an illusory nature of that group due to, e.g., filamentary structures projected onto a small area (Hernquist, Katz & Weinberg 1994). In such case, even if an intra-filament medium existed, galaxies and gas would have modest relative velocity.

We stress that *none* of the numerous effects described above are expected to be important in a low–density universe. In this case most of the group mass would have been assembled at relatively high redshifts, with little or no secondary infall. This scenario would be more similar to that previously studied (e.g. Barnes 1985; Diaferio, Geller & Ramella 1994), i.e., an expanding top-hat overdensity that collapses and evolves in isolation, with strong merging on a time scale of a few CTs controlling the overall morphology of the group.

## 5. Conclusions

This work concerns results and interpretation of N–body simulations of small groups of galaxies in an Einstein–de Sitter universe. Our models describe formation and evolution of groups in terms of the collapse of a spherical overdense region followed by secondary infall of the surrounding mass. We find that merging is effective at redshifts $z \approx 1 - 0.35$, but its rate decreases and becomes negligible for the subsequently infalling galaxies. This is understood on the basis of two mechanisms, both linked to the presence of substantial secondary infall. First, efficient tidal stripping during the first close passage through the center of the group for late infalling galaxies reduces their mass and geometrical cross section. Second, the orbital decay time due to dynamical friction is long for the same infalling galaxies.

We then used the PS formalism, recast in terms of a rate equation, to exhibit the difference between high and low–density universes as for the age of such structures. A larger percentage of recent groups, both in terms of absolute age and in terms of internal CTs, is found in the case of a critical universe where an efficient turnover rejuvenates the population even for $z < 0.3$.

We looked for compact groups and their characteristic in our set of simulations. We conclude that recent formation and membership replenishment by infalling galaxies concur to make compact groups frequent structures in a high–density universe, independently to a large extent of the particular initial power spectrum chosen. Specifically, our secondary infall model is consistent with the abundance of Hickson compact groups, and leads us to argue that the existence today of many small, dense galaxy associations like HCGs points toward a high–density universe. In addition, the secondary infall picture gives a likely explanation of some apparent puzzles related to CGs, like: lack of signatures of recent merging in ellipticals; dynamical signs of interactions, and gas deficiency in spirals.

Our model is able to give both a likely description of the dynamical processes associated to the dense groups environment and to predict a number density consistent with observations. While Hickson catalog is likely to include different kinds of associations (mere superpositions, CG inside loose groups, filaments, etc.), many truly compact, bound configurations are being confirmed by the very presence of signs of galaxy interactions, and more directly by X–ray diffuse emission.

This study predicts that a real compact group in a critical universe will appear as an association of galaxies comprising at least one bright elliptical, without obvious signs of recent merging activity, plus a number of later type galaxies with signs of weak interactions.

We stress that real CGs and more generally small groups, constitute an important means to investigate $\Omega_0$. It would be of great interest to test the prediction of our simple scheme with high resolutions simulations of group formation in flat cosmologies with non–zero cosmological constant, where we expect dynamical features close to those in a critical universe (Tozzi, Cavaliere & Governato 1995). Moreover observations of spirals in CGs aimed at testing the hypothesis of their secondary infall origin would share light on the dynamical processes involved in the formation of small groups of galaxies.

We thank Nicola Menci, Claudia Mendes de Oliveira, Julio Navarro and Jack Sulentic for insightful comments, and Lars Hernquist and Julio Navarro for providing their N-body codes. We also thank Isabella Prandoni and Angela Iovino for giving us part of the software necessary for the CGs search algorithm. We are grateful to a referee for comments that helped us toward improving the paper. FG thanks Mick, Keith, Charlie, Ron & Bill for their help during the writing of this paper.

| Run | $\delta_i$[a] | % $M_{gal}$[b] | $z_{coll}$[c] | $z_{ta}$[d] | %infall[e] |
|---|---|---|---|---|---|
| r108g10 | 1.08 | 10 | 1.2 | 2.5 | 50% |
| r108g20 | 1.08 | 20 | 1.2 | 2.5 | 50% |
| r108g40 | 1.08 | 40 | 1.2 | 2.5 | 50% |
| r71g10 | 0.71 | 10 | 0.68 | 1.6 | 37.5% |
| r71g20 | 0.71 | 20 | 0.68 | 1.6 | 37.5% |
| r71g40 | 0.71 | 40 | 0.68 | 1.6 | 37.5% |
| r53g10 | 0.53 | 10 | 0.37 | 1.15 | 25 % |
| r53g20 | 0.53 | 20 | 0.37 | 1.15 | 25 % |
| r53g40 | 0.53 | 40 | 0.37 | 1.15 | 25 % |
| r35g10 | 0.35 | 10 | 0. | 0.6 | 0% |
| r35g20 | 0.35 | 20 | 0. | 0.6 | 0% |
| r35g40 | 0.35 | 40 | 0. | 0.6 | 0% |
| r108g20s | 1.08 | 20 | 1.2 | 2.5 | 50% |
| r71g20s[f] | 0.71 | 20 | 0.68 | 1.6 | 37.5% |
| r71g10-2[f] | 0.71 | 10 | 0.68 | 1.6 | 37.5% |
| r71g40-2[f] | 0.71 | 40 | 0.68 | 1.6 | 37.5% |

Table 1: Parameters of the simulations

[a] initial overdensity of the central region
[b] mass fraction in galaxies
[c] redshift of collapse of the overdense region
[d] redshift of turn-around of the overdense region
[e] percentage of infalling mass
[f] same random seed

| Galaxy model | $r_{50}$ STARS[a] | $r_{50}$ DM [b] | $r_{90}$ STARS+DM[c] |
|---|---|---|---|
| g10 | 0.2 | 0.6 | 1.2 |
| g20 | 0.2 | 1.6 | 2.8 |
| g40 | 0.2 | 3.6 | 6.2 |

Table 2: Galaxy models

[a] 3D Half-mass radius of STARS
[b] 3D Half-mass radius of DM
[c] 3D 90% mass radius (STARS+DM)

| Overdensity value | CT[a] | # of CTs since $z_{coll}$ [b] |
|---|---|---|
| 1.08 | 8 | 6.5 |
| 0.71 | 12 | 3.4 |
| 0.53 | 16 | 1.75 |
| 0.35 | 26 | 0. |

Table 3: CT of overdense regions

---

[a] CT in time units
[b] Number of CT from redshift of collapse and present time

Fig. 1.— Snapshots of simulation r71g2-2 at $z = 1.5(a), 0.5(b), 0(c)$ (only STARS and DM halo particles are plotted). Note the change of scale from panels a and b to c.

Fig. 2.— Panel a: Number of surviving galaxies vs. time for runs r71g10-2 (continuous line), r71g20s (dotted) and r71g40-2 (dashed); these runs used the same random seed for generating the initial conditions. Panel b: same for runs r53; continuous line: model g10, dotted : g20, dashed: g40.

Fig. 3.— Bound mass (DM+STARS) for some infalling galaxies in different runs. Continuous line: model g10; dotted: g20; dashed: g40.

Fig. 4.— $T_{fr}$ as a function of the turn–around radius $R_{ta}$ is shown for different values of $J/J_c$ and $M_g$, and for the interesting range of turn–around radii. The horizontal line corresponds to the time span covered by the simulations.

Fig. 5.— The number density of collapsed objects in the mass range of groups is plotted, as given by eq. 3.1 for the two models in the text and for the intermediate spectral index value $n = -2$. Solid line: $\Omega_0 = 1$, $b = 1.7$; dotted line: $\Omega_0 = 0.2$, $b = 1$.

Fig. 6.— Formation and destruction rates vs. $z$ for the power law spectrum with $n = -2$, for the two models $\Omega_0 = 1$, $b = 1.7$ (solid line); and $\Omega_0 = 0.2$, $b = 1$ (dotted line).

Fig. 7.— Probability vs. time to observe a HCG configuration in four different runs. The time interval corresponds to $z = 0.1 - -0$.